# Low energy range dielectronic recombination of Fluorine-like Fe$^{17+}$ at the CSRm[*]

Nadir Khan[1,2], Zhong-Kui Huang(黄忠魁)[1], Wei-Qiang Wen(汶伟强)[1†], Sultan Mahmood[1], Li-Jun Dou(豆丽君)[1,2], Shu-Xing Wang(汪书兴)[3], Xin Xu(许鑫)[3], Han-Bing Wang(汪寒冰)[1], Chong-Yang Chen (陈重阳)[4], Xiao-Ya Chuai(啜晓亚)[1,2], Xiao-Long Zhu(朱小龙)[1], Dong-Mei Zhao(赵冬梅)[1], Li-Jun Mao(冒立军)[1], Jie Li(李杰)[1], Da-yu Yin (殷达钰)[1], Jian-Cheng Yang(杨建成)[1], You-Jin Yuan(原有进)[1], Lin-Fan Zhu(朱林繁)[3] and Xin-Wen Ma(马新文)[1‡]

[1]Institute of Modern Physics, Chinese Academy of Sciences, Lanzhou 730000, China
[2]University of Chinese Academy of Sciences, Beijing 100049, China.
[3]Hefei National Laboratory for Physical Sciences at Micro scale, Department of Modern Physics, University of Science and Technology of China, Hefei 230026, China
[4]Institute of Modern Physics, Fudan University, 200433, Shanghai, China

**Abstract:** The accuracy of dielectronic recombination (DR) data for astrophysics related ions plays a key role in astrophysical plasma modeling. The measurement of the absolute DR rate coefficient of *Fe$^{17+}$* ions was performed at the main cooler storage ring at Institute of Modern Physics, Lanzhou, China. The experimental electron-ion collision energy range covers first Rydberg series up to *n* = 24 for the DR resonances associated with the $^2P_{1/2} \rightarrow {}^2P_{3/2}$ Δ*n* = 0 core excitations. A theoretical calculation was performed by using FAC code and compared with the measured DR rate coefficient. Overall reasonable agreement was found between the experimental results and calculations. Moreover, plasma rate coefficient was deduced from the experimental DR rate coefficient and compared with the available results from the literature. At the low energy range significant discrepancies were found therein and the measured resonances challenged state-of-the-art theory at the low collision energies.

## 1 Introduction

Astrophysical plasmas can be divided into two broad classes, photoionized plasma and collisionally ionized plasma [1]. Photoionized plasma forms in the media surrounding the cosmic sources such as active galactic nuclei (AGN), cataclysmic variable stars and X-ray binaries, where the ionization is because of photon [2]. However, the collisionally ionized plasma is mostly found in solar coronae, supernova remnant, galaxies and in intercluster medium in clusters of galaxies, where the ionization is by electron impact [3]. In order to understand the properties of the astrophysical plasmas, the new generation X-ray observatories, such as *ASCA* [4], *Chandra* (NASA) [5] and *XMM-Newton* (ESA)[6], have been launched to observe the high resolution X-ray spectra from various cosmic sources. All the observed spectra have to be interpreted by plasma modelling. However, most of the input atomic data for the plasma modeling are from theory. For electron-ion collision processes in astrophysical plasmas, dielectronic recombination (DR) is one of the important recombination process, which determines the charge state distribution and ionization balance therein. Therefore, precise DR rate coefficients are an issue of major concern for astrophysical plasma modeling [7-10]. In case of collisionally ionized plasma, theoretical DR data are now available in literature [3, 11, 12] with rather good agreement to the experimental data for plasma modeling [13]. To use this available atomic data for X-ray astrophysics implications, the atomic databases such as XSPEC [14], AtomDB version 1.3 and AtomDB version 2.0 [15] are widely used to model the astrophysical plasma. However, for DR in the photoionized plasma which located in the low-energy range, the modeling is mostly based on theoretical predictions and calculations，and many theories cannot provide the sufficiently precise DR rate coefficients [16]. In addition, the recent experimental approach for low energy range DR investigation has also shown that earlier computations of low temperature DR rate coefficients are not accurate [17-19]. To model the line emission, thermal and ionization structures of plasmas, the astrophysicists required accurate bench-mark atomic data from electron-ion recombination experiments [20].

[*] Project supported by the National Key R&D Program of China under Grant No. 2017YFA0402300, the National Natural Science Foundation of China through No. 11320101003, No.U1732133, No. 11611530684 and Key Research Program of Frontier Sciences, CAS, Grant No. QYZDY-SSW-SLH006.
[†] Corresponding author. E-mail: wenweiqiang@impcas.ac.cn
[‡] Corresponding author. E-mail: x.ma@impcas.ac.cn

Iron is the most abundant heavy element in astrophysical plasmas and has a great importance in astrophysics [9, 21]. The high resolution X-ray spectra from ~14 Å to ~17 Å were observed from different active galactic nuclei (AGN) such as luminous quasar IRAS 13349+2438 [22], Seyfert 1 galaxy NGC-3783 [23]. The rich absorption features contribution of iron ions have been seen in these spectra , when analyzed by using photoionization codes CLOUDY [24] and XSTAR [25]. However large discrepancies were found between observed spectra and the results obtained from available DR theoretical data for iron ions. These discrepancies were due to under estimation of low-temperature DR rate coefficient by available models for L-shell and M-shell iron ions. In order to solve this problem, electron-ion recombination experiments on different charge state of iron ions have been initiated at the test storage ring (TSR) [7, 26], Heidelberg Germany. The purpose was to provide the accurate experimental DR data and reduce the uncertainties in calculations. For the case of F-like iron ion, most of the earlier calculations neglect the contribution from the fine structure $2p_{3/2}-2p_{1/2}$ excitations which have been shown to be very important for low-temperature DR rate coefficient [27]. Especially for photoionized plasma modelling the inclusion of fine-structure excitation is very important for producing the reliable DR rate coefficient [26, 28, 29]. The other important astrophysical aspect of the fluorine-like ions forming neon-like ions is the determination of solar and stellar upper atmosphere abundances [30]. Here, we present absolute electron-ion recombination rate coefficients of fluorine-like $Fe^{17+}$ from an experiment at the main cooler storage ring (CSRm) and also from a theoretical calculation using flexible atomic code (FAC) [31]. It should be noted that the electron-ion merged beams technique at the heavy-ion cooler storage rings is the only laboratory method capable of studying DR at low collision energy, it also provides a high resolution with low background measurement of DR for precision atomic spectroscopy [19, 32, 33].

Dielectronic recombination is a two-step process, where one free electron captures in one of Rydberg states of the ion with simultaneous excitation of a core electron and produces doubly excited intermediate state. This process completes when the system stabilizes itself to below ionization threshold by emitting excess energy in form of photon. Another co-existed recombination process called radiative recombination also occurs at the same time. RR is the process where one free electron is captured into a bound state of the ion and a photon is emitted. For electron-ion recombination of F-like $Fe^{17+}$, RR can be expressed as

$$Fe^{17+} + e^- \rightarrow Fe^{16+} + h\nu \quad (1)$$

and DR for $\Delta n = 0$ transitions can be written as

$$Fe^{17+}(2s^2 2p^5[^2P_{3/2}]) + e^-$$
$$\rightarrow \begin{cases} Fe^{16+}(2s^2 2p^5[^2P_{1/2}]\,nl\,)\ n=18,19...,\infty \\ Fe^{16+}(2s 2p^6[^2S_{1/2}]\,nl\,)\ n=6,7,...,\infty \end{cases} \quad (2)$$

In the present recombination experiment of F-like iron, the experimental electron-ion collision energy range was $0-6\ eV$ in the center of mass frame (c.m). It covers first Rydberg series associated with the transition $^2P_{3/2} \rightarrow\ ^2P_{1/2}$, where $n$ is principal quantum number and can be resolved up to $n = 24$.

The paper is organized as follows: Section 2 gives a brief introduction to the experimental method and data analysis. The experimental, calculated as well as plasma rate coefficients are presented and discussed in section 3. A conclusion is given in section 4.

## 2 Experimental method

The experiment was performed at the main cooler storage ring (CSRm), at Institute of Modern Physics Lanzhou, China [34]. The details about DR experiments at the CSRm were well described in references [18, 35]. Here we just briefly describe the DR experiment of $^{56}Fe^{17+}$. The F-like $Fe^{17+}$ ions were produced in the super-conducting Electron Cyclotron Resonance (ECR) ion sources, and accelerated in Sector Focused Cyclotron (SFC) up to an energy of $E_{ion}$ = 6.08 $MeV/u$. After that, the ion beam was injected into the CSRm and stored in the ring. The storage lifetime of the ion beam was around 20 $s$. The beam current was $I_{ion}$ ~ 350$\mu A$, corresponding to 2.3 ×10$^8$ ions. The electron-cooler was employed to cool the ion beams and also used as an electron-target for electron-ion recombination experiment. The electron beam was produced at the cathode and collected at the anode of the 35 kV electron cooler (EC-35). The magnetic fields applied at the cathode and cooler section were 1250 *Gs* and 390 *Gs,* respectively, that allow adiabatic expansion of the electron beam with expanded diameter of $d \sim 50$ *mm*.

The circulating ion beam merged with electron beam to an effective length of 4 *m* in the electron cooler EC-35 at the CSRm. The mean velocity of electron beam matched to the mean velocity of ion beam at cooling point. The detuning voltage $U_d$ was applied to the cathode of the electron cooler to change the electron's kinetic energy relative to the ions according to a specific time scheme *i.e.* 10 *ms* detuning and 190 *ms* of cooling [36]. The recombined ion beam was separated from the primary ion beam in the first dipole magnet downstream the electron cooler. Finally, the recombined $Fe^{16+}$ ions were detected by a scintillator detector (YAP: Ce + PMT) with ~ 100% efficiency [37]. During the whole measurement, a Schottky pick-up system was used to monitor the revolution frequency and longitudinal momentum spread of the ion beam. The momentum spread of the ion beam was deduced about $\Delta p / p \sim 3.4 \times 10^{-4}$ from the Schottky spectrum.

Data acquisition was started after 3 seconds of electron cooling following the beam injection. The electron-ion recombination rate coefficient can be determined from

$$\alpha(E) = \frac{R}{N_i n_e (1 - \beta_e \beta_i)} \cdot \frac{C}{L} \quad (3)$$

where, R is the count rate, $N_i$ is the number of stored ions, $n_e$ is the density of electron beam, $\beta_e$ and $\beta_i$ are the velocities of the electrons and ions respectively, C is the circumference of the ring about 161.00 *m* and L is length of interaction region [35]. In order to obtain the recombination rate coefficient, the electron-ion collision energy from laboratory frame system has to be transformed to center-of-mass-frame (c.m.) system. To calculate the relative collision energy (*i.e.* $E_{rel}$) between electrons and ions in the c.m. system, the following relativistic formula was used

$$E_{rel} = \sqrt{m_e^2 c^4 + m_i^2 c^4 + 2 m_e m_i \gamma_e \gamma_i c^4 (1 - \beta_e \beta_i \cos\theta)} - m_e c^2 - m_i c^2 \quad (4)$$

where $m_i$ and $m_e$ represent masses of ion and electron. $\gamma_i$, $\gamma_e$ and $\beta_i$, $\beta_e$ are Lorentz factors and relativistic factors for ion beam and electron beam, respectively. *c* is the speed of light and $\theta$ is the angle between ion and electron beams, which was always optimized less than 0.1 mrad during measurement.

The space charge effect of the electron-beam was taken into account for calculating the relative collision energy, as the effective electron beam energy ($E_e$) is

$$E_e = -e(U_{cath} + U_d + U_{sp}) \quad (5)$$

where $U_d$ is referred as detuning voltage and $U_{sp}$ is the space charge potential. The space-charge potential is modelled by the formula

$$U_{sp}(v_e) = (1-\zeta) \frac{I_e r_c m_e c^2}{v_e e^2} \left[ 1 + 2\ln\left(\frac{b}{a}\right) - \left(\frac{r}{a}\right)^2 \right] \quad (6)$$

Here $I_e$ is the electron beam current, $r_c$ the classical electron radius, $m_e$ is the electron rest mass, $c$ is the speed of light, $v_e$ is the electron velocity, $e$ is the elementary charge, $r$ is the distance from the electron beam axis, $a = 2.8\ cm$ and $b = 20\ cm$ are the radii of the electron beam and the cooler tube, respectively. The parameter $\zeta$ accounts for the residual gas ions that usually trapped in the electron beam. The calculated space-charge potential from experimental parameters at cooling point was $U_{sp} \sim 140$ V.

## 3 Results and discussion

### 3.1 Electron-ion recombination rate coefficient

Figure 1 shows the electron-ion recombination rate coefficient as a function of electron-ion collision energy for F-like $Fe^{17+}$ ions. The series associated to the $\Delta n = 0$ core excitations from $2s^22p^5(^2P_{3/2})\ nl$ to $2s^22p^5(^2P_{1/2})\ nl$ were observed. The resonance positions were obtained from Rydberg formula

$$E_{res} = \Delta E - \text{Ry}\left(\frac{q}{n}\right)^2 \tag{7}$$

where Ry = 13.606 $eV$ is Rydberg constant, $q = 17$ is the charge state of ion, $\Delta E = 12.7182$ $eV$ is the core excitation energy taken from NIST database [38]. The first Rydberg series of intermediate $Fe^{16+}$ resonant states are identified from $n=18$ up to $n = 24$.

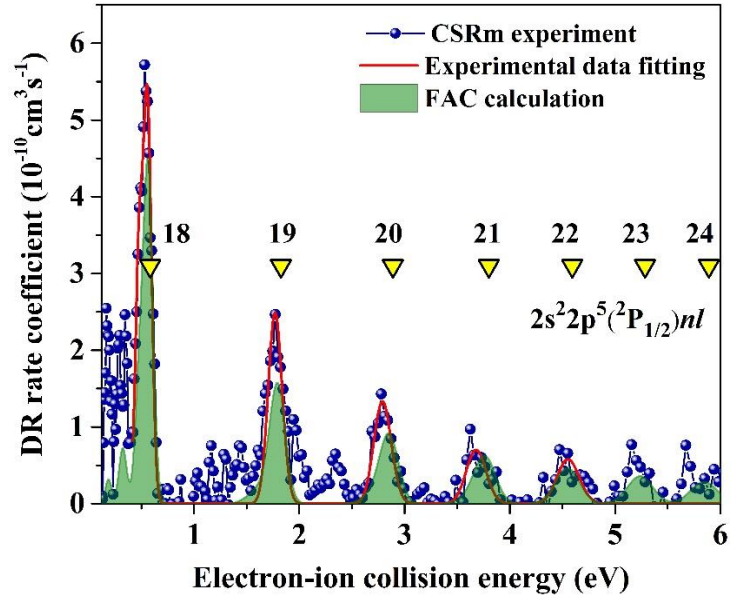

Fig. 1. Dielectronic recombination rate coefficient for $Fe^{17+}$ ions from measurement (connected blue dots) are compared with a calculation based on FAC code (shaded green area). The yellow triangle indicate calculated Rydberg states associated to $^2P_{1/2} \to {}^2P_{3/2}$ core transitions. The solid red line shows the fitting result for obtaining temperature of measured DR rate coefficient at CSRm.

It should be noted that the recombination rate coefficient are obtained by convolution of resonance cross sections $\sigma_d$ ($v$) with an asymmetrical Maxwell-Boltzmann distribution function, resembling the distribution of relative velocities between the electrons and the circulating ions, *i.e.*

$$\alpha(E_{rel}) = \int \sigma(v) v f(\vec{v}, \vec{v}_{rel}) d^3\vec{v} \quad (8)$$

where, $\sigma(v)$ is the energy-averaged cross section of the DR process. The theoretical cross section of state $d$ is written as

$$\hat{\sigma}_d(v) = \frac{2\pi\hbar Ry}{E_d} \pi a_0^2 \frac{g_d}{2g_i} \frac{A_a(d \to i) \sum_f A_r(d \to f)}{\sum_k A_a(d \to k) + \sum_{f'} A_r(d \to f')} \quad (9)$$

Where $\hat{\sigma}_d(v)$ is known as the strength of the resonance state $d$ and is defined as energy integrated cross section, $Ry$ is the Rydberg constant, $E_d$ is the resonance energy, $a_0$ is the Bohr radius. $g_i$ and $g_d$ are the statistical weights of the initial ionic core and of the intermediate states, $A_a$ and $A_r$ are autoionization and radiative decay rates, respectively. In summations $k$ denotes all the states which are attainable by autoionization of the intermediate state and $f$ runs over all states below the first ionization threshold, $f'$ includes all states below $d$ [31, 32]. In Eq. (8), the $f(\vec{v}, \vec{v}_{rel})$ is a flattened Maxwellian distribution function of the electron beam and expressed by

$$f(\vec{v}, \vec{v}_{rel}) = \frac{m_e}{2\pi k_B T_\perp} \exp\left(-\frac{m_e v_\perp^2}{2k_B T_\perp}\right) \times \left[\frac{m_e}{2\pi k_B T_\parallel}\right]^{1/2} \exp\left(-\frac{m_e(v_\parallel - v_{rel})^2}{2\pi k_B T_\parallel}\right) \quad (10)$$

where $m_e$ is the mass of electron, $k_B$ is the Boltzmann constant, $T_\perp$ and $T_\parallel$ are experimental electron velocity distribution of the parallel and perpendicular temperatures with respect to the electron beam propagation direction. $v_{rel}$ is the relative velocity between electron and ion [39].

By fitting the measured DR spectrum from 0.38 *eV* up to 5 *eV* with six resonances associated to $2s^2 2p^5$ ($^2P_{1/2}$) *nl*, where $n$ = 18, 19, 20, 21, 22. The resonance energies and strengths are obtained and listed in Table 1. The electron beam temperatures obtained from this fitting are $K_B T_\parallel = 1.2$ *meV* and $K_B T_\perp = 11$ *meV*. The energy resolution was achieved less than $\Delta E \sim 0.09$ *eV* at full width at half maximum (FWHM) around $E_{rel} \sim 0.49$ *eV*.

The uncertainty of the experimental recombination rate coefficients is estimated to be about 30%, an uncertainty of 10% due to combination of counting statistics, electron and ion beam currents, and interaction length, and an uncertainty of 20% due to the electron density distribution profile and also the position of the ion beam in this profile.

Table 1. Resonance energies and strengths from the fitted DR resonances of the measured recombination rate coefficient from 0.18 to 5 *eV* (See graph fitting in Fig. 1). Number in parenthesis represents the uncertainty.

| $E_d$ (eV)   | $\sigma_d$ ($10^{-21} cm^2 eV$) |
|--------------|-------------------------------|
| 0.19 (0.03)  | 588 (39)                      |
| 0.33 (0.03)  | 517 (39)                      |
| 0.49 (0.04)  | 704 (54)                      |
| 0.57 (0.04)  | 1093 (54)                     |
| 1.79 (0.07)  | 521 (38)                      |
| 2.81 (0.09)  | 278 (38)                      |
| 3.70 (0.10)  | 146 (39)                      |
| 4.57 (0.11)  | 121 (39)                      |

The theoretical calculations were performed by using FAC code [40]. The doubly excited states $2s^2 2p^5 [^2P_{1/2}] nl$, $n = 18 \sim 24$ of Ne-like Fe$^{16+}$ ions were included. Here $l$ value was included up to $l_{max} = 20$. For the Fe$^{17+}$ ions, all possible electronic-dipole transitions from the $2s^2 2p^5 [^2P_{1/2}] nl$ resonances were considered. The theoretical rate coefficients were obtained by convoluting the calculated resonance cross sections with the experimental electron energy distribution (see Eq. (8)). The calculated DR rate coefficients are shown by green area in Fig. 1. It can be found that the measured rate coefficients and the theoretical calculation are in reasonable good agreement. However, the theory is slightly lower than the experiment at low energy range. Which means that the calculated result by FAC code could not reproduce DR rate coefficient both in energy positions and intensities comparable to the measured results at very low energy range. These discrepancy associated to the DR resonances $Fe^{16+}$ $2s^2 2p^5$ $(^2P_{1/2})nl$, where $n=18\sim24$.

The $n$-sum resonance strengths derived from the experimental data for energy range of 0.38 $eV$−5 $eV$ are compared with the experimental data from TSR storage ring [13] as shown in Table 2. In addition, the calculated strengths by FAC code (this work), and the previously calculated results by state-of-the-art codes Multi-Configuration Dirac Fock (MCDF) and multi-configuration Breit-Pauli (MCBP) are also shown in Table. 2. Because of statistical uncertainty in experimental measurement below 0.38 $eV$ the first two peaks related to 18$s$ and 18$p$ were not measured accurately, so $n$ = 18$l$ value was only considered for 18$d$ and 18$l$ with $l \geq 3$ peaks for all the data under the comparison. A good agreement between our data and TSR data can be found for the resonance strengths of $n$ = 18~22. This comparison shows that CSRm can also provide reliable experimental data to benchmark theory for astrophysical plasma modeling and for precision spectroscopic investigation. However, clear difference between the measured and calculated DR resonance strengths and also little difference between different theoretical calculations can be found in Table. 2. As a result, further precisely experimental results and also theories of the DR rate coefficients for highly charged ions are required.

Table 2. Resonance strengths of first five DR resonances from this work (measurement at CSRm and FAC calculations) and from previous work (measurement at TSR and MCDF, MCBP calculations) for $\Delta n = 0$ [13, 41]. Here $\sigma_d$ represents the energy-integrated cross sections for $2s^2 2p^5(^2P_{1/2})nl$ resonances.

| n | $\hat{\sigma}_d$ ($10^{-21}$ cm² eV) | | | | |
|---|---|---|---|---|---|
| | CSRm (Experiment) | TSR (Experiment) | FAC (This work) | MCDF | MCBP |
| 18 | 1797 (77) | 2018 (13) | 1417.5 | 1557.4 | 1634.5 |
| 19 | 521 (38) | 606 (14) | 427.1 | 449.7 | 477.6 |
| 20 | 278 (38) | 336.5 (8.6) | 227.4 | 239.7 | 252.2 |
| 21 | 146 (39) | 205.4 (6.9) | 149.1 | 154.5 | 161.1 |
| 22 | 121 (39) | 140.7 (4.2) | 105.5 | 111.1 | 113.2 |

Note: In this table the strengths for sum of $n = 18l$ value does not include the contribution from $18s$ and $18p$. For more details see text.

### 3.2 Plasma rate coefficient

The plasma rate coefficients, which is useful for astrophysical plasma modelling, can be obtained by convoluting recombination rate coefficient with the Maxwell-Boltzmann energy distribution of the electrons in a plasma as [42, 43]

$$\alpha(T_e) = \int \alpha(E) f(E, T_e) dE \qquad (11)$$

where, the term $\alpha(E)$ represents the measured electron-ion recombination rate coefficient and $f(E, T_e)$ is the average Maxwellian temperature distribution function as given by

$$f(E, T_e) = \frac{2E^{1/2}}{\pi^{1/2}(k_B T_e)^{3/2}} \exp\left(-\frac{E}{k_B T_e}\right) \qquad (12)$$

where, $E$ is the relative energy and $T_e$ is the electron temperature.

As shown in Fig. 2, the plasma rate coefficients for DR of $Fe^{17+}$ were deduced from the measured electron-ion recombination rate coefficients in the temperature range from 0.1 $eV$ to 4.7 $eV$. The values of strengths and energy positions were used to obtain the plasma rate coefficient, which were extracted from fitting of DR rate coefficient and compared with the calculated data from FAC code and also from the literature. The plasma rate coefficients derived from the measurement at the CSRm and FAC calculation are shown by thick blue solid and thin green solid lines, respectively. The previous results from the measurement at the TSR are indicated by dash-dot line, and the corresponding theoretical calculations by using MCBP and MCDF are denoted by red dashed curve and black dotted curve, respectively. At the temperature from 0.1 $eV$ to 1.0 $eV$ the theoretical calculation from FAC are ~ 30% lower than the CSRm experimental results. At this low energy range the discrepancies can be also interpreted as from change of resonance positions, because the plasma rate coefficient is very sensitive to changes in resonance positions and strengths in merged beam recombination rate coefficient at low energy range. A small change in position and strengths translate into large discrepancies in plasma rate coefficient [17]. However, at temperature range from 2.0 eV to 4.7 eV, a very good agreement was found between experimental results and FAC calculation. The plasma rate coefficient from CSRm agrees very well with the TSR data from 0.1 eV to 0.3 eV, and is about ~20% lower than the TSR data from 0.5 eV to 4.7 eV. It can be found that MCDF and MCBP results underestimate the plasma rate coefficient at this temperature range, and the clear discrepancies can be seen in plasma rate coefficient measured by TSR (dash-dot line) and the all the theoretical calculations. In the temperature range of 0.2 $eV$ to 4.7 $eV$ all the calculations underestimate the plasma rate coefficient of about 30% as compared with the TSR. These discrepancies comprises the fact that calculation of accurate DR resonance structure at low energy collisions is still a very challenging task even for state-of-the-art codes.

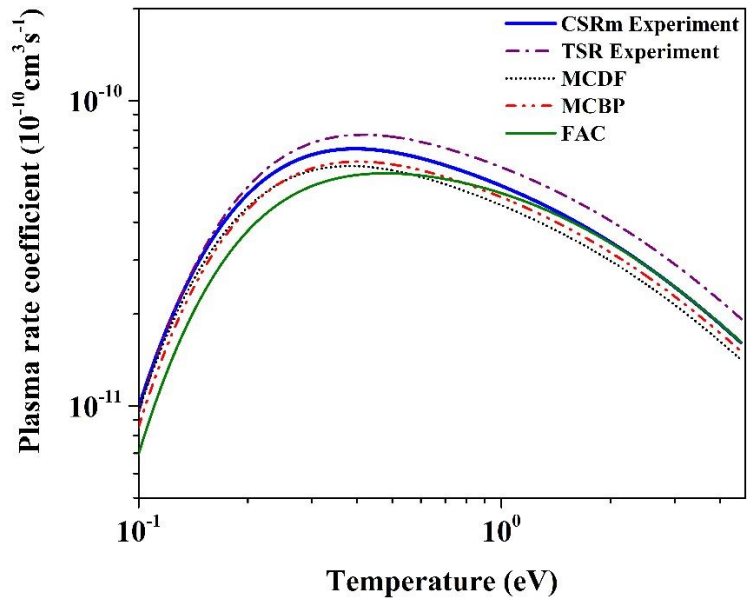

Fig. 2. Comparison of plasma rate coefficients derived from the experimental result with the calculated results from FAC code and also the existed plasma rates coefficients from literature. The plasma rate is derived from 18~22 resonances strengths and the contribution from 18$s$ and 18$p$ are not take into account as indicated in Table 2. The thick solid blue line denote experimental results from CSRm and thin solid green line represents FAC calculation. The experimental result from TSR is displayed by Purple dash-dot line and corresponding calculations by MCBP and MCDF are shown by red dashed curve and black dotted curve, respectively. Since our measurement energy range is only up to 5 eV and the contribution from 18$s$ and 18$p$ are missing in this figure, this plasma rate cannot be used in plasma modeling.

## 4 Conclusions

The DR rate coefficient of F-like iron in energy range 0−6 $eV$ have been measured by employing the electron-ion merged beams method at the CSRm at Lanzhou, China. The measured energy range covers the first Rydberg series of $^2P_{1/2}$ to $^2P_{3/2}$ core transitions of $\Delta n = 0$ up to $n = 24$. A FAC code was employed to calculate the DR rate coefficient to compare with the measured results. A reasonable good agreement between the experimental results and the calculations could be found by taking into account of the estimated 30% experimental uncertainty. The plasma rate coefficient derived from the electron-ion recombination rate coefficient was compared with the FAC calculation and also the available data in literature, and overall a reasonable agreement was found. However, the discrepancies between experimental and theoretical results can be seen at low temperature range which can be mainly attributed to the limited accuracy of the theoretical calculation. Our measurement challenges modern DR theory to calculate accurate electron-ion recombination rate coefficient of multi-electron ions at low electron-ion collision energies.

*Nadir Khan acknowledges China Scholarship Council (CSC) of China for providing opportunity and fund for him to study at IMP, UCAS. W. Wen thanks the support by the Youth Innovation Promotion Association CAS. We would like to say thanks to Daniel Wolf Savin for providing their data on internet and Stefan Schippers for guiding us in the data analysis. The authors would like to thank the crew of Accelerator Department for skillful operation of the CSR accelerator complex.*